\documentclass[a4paper,11pt]{article}
\usepackage{pos}
\usepackage{listings}
\usepackage[title]{appendix}
\usepackage{floatrow}
\usepackage{url}
\usepackage{hyperref}
\newfloatcommand{capbtabbox}{table}[][\FBwidth]

\usepackage{xcolor}
\definecolor{codegreen}{rgb}{0,0.6,0}
\definecolor{codegray}{rgb}{0.5,0.5,0.5}
\definecolor{codepurple}{rgb}{0.58,0,0.82}
\definecolor{backcolour}{rgb}{0.95,0.95,0.92}

\lstdefinestyle{mystyle}{
    backgroundcolor=\color{backcolour},   
    commentstyle=\color{codegreen},
    keywordstyle=\color{magenta},
    numberstyle=\tiny\color{codegray},
    stringstyle=\color{codepurple},
    basicstyle=\ttfamily\footnotesize,
    breakatwhitespace=false,         
    breaklines=true,                 
    captionpos=b,                    
    keepspaces=true,                 
    numbers=left,                    
    numbersep=5pt,                  
    showspaces=false,                
    showstringspaces=false,
    showtabs=false,                  
    tabsize=2
}

\title{Maximizing the Bang Per Bit}

\author*[a]{M. A. Clark}
\author[b]{Dean Howarth}
\author[a]{Jiqun Tu}
\author[a]{Mathias Wagner}
\author[a]{Evan Weinberg}

\affiliation[a]{NVIDIA Corporation,\\
  2788 San Tomas Expressway Santa Clara, CA 95051, United States of America}
\affiliation[b]{Lawrence Berkeley National Laboratory,\\
  1 Cyclotron Rd, Berkeley, CA 94720, United States of America}

\emailAdd{mclark@nvidia.com}

\abstract{Reducing memory traffic is critical to accelerate Lattice QCD computations on modern processors, given that such computations are memory-bandwidth bound.  A commonly used strategy is mixed-precision solvers, however, these require careful treatment to ensure stable convergence.  We give an overview of the strategies employed in QUDA to stabilize mixed-precision variants of Conjugate Gradient (CG), and its multi-shift brethren.  Through the use of customized numerical storage formats we can significantly improve upon the precision achievable compared to IEEE numerical formats, increasing both the solver precision and stability achievable at fixed word size. We give examples using BiCGStab(\(l\)) and multi-shift CG solvers using the HISQ operator.
}

\FullConference{%
  The 39th International Symposium on Lattice Field Theory (Lattice2022),\\
  8-13 August, 2022 \\
  Bonn, Germany
}

\begin{document}
\maketitle

\section{Introduction}
\vspace{-2mm}

It is well known that lattice quantum chromodynamics (LQCD)
calculations are memory bandwidth bound on modern computing architectures,
this stemming from the low arithmetic intensity of the discretized
fermion operators (stencils) which are the computational dominant kernel in
most LQCD computations.  Given that the expected future
trajectory of computer architecture will exacerbate the
gap between computational throughput and memory bandwidth, we cannot
expect future supercomputers to remedy this situation.

In order to soften this bottleneck there have been a number of
techniques deployed, both as part of general HPC developments, and in
a domain-specific approaches specifically for LQCD
computations.
\begin{itemize}
\item Cache blocking: the number of redundant loads can be minimized
  by utilizing a cache tiling-strategy, where temporal locality is exorcised to reuse loads in on-chip cache~\cite{6114421, 6339591}. 
\item Block solvers: by deploying a batched linear solver, where
  multiple right hand sides are solved simultaneously, the memory
  traffic to load the stencil coefficients can be amortized.  This
  transformation effectively turns a matrix-vector problem, into a
  matrix-matrix problem, increasing the arithmetic intensity as a result~\cite{Clark:2017ekr}.
\item Compression: LQCD is a theory that is blessed with many
  symmetries that can be harnessed to reduce their memory footprint.  In particular, the link-field matrices that appear in the Dirac stencil are, depending on the smearing used,
  usually a member of the SU(3) group.  This allows one to reduce the
  real numbers required to exactly store each matrix from 18 to 12 or as few
  as eight real numbers~\cite{Egri_2007, Clark:2009wm}.
\item Mixed-precision Krylov solvers: using less precision to store fields
  means less time spent loading / storing those field.  The most
  commonly used approach is to use ``double-single'' solvers, where
  the bulk of the work is done using IEEE 754  single precision, {\it FP32},  with correction to IEEE 754 double precision, {\it FP64}, though ``double-half'' is also
  deployed, particularly in the QUDA library (see \(\S\ref{sec:ieee}\)).
\end{itemize}

In this work we focus on a new approach to tackle the memory bandwidth
crunch: {\bf how to improve precision at a fixed number of bits}.
In doing so we ask ourselves, what would the optimum
numerical format be for an LQCD-type problem?  Almost all modern
microprocessors that are capable of floating point adopt the IEEE 754
convention, however there is no
reason why the IEEE formats are the optimum ones for LQCD data.

The fact that we are constrained to using IEEE floating point for the
actual computation is not an issue since we are free to decouple the
computation format from the storage format, with the expectation that
any custom format would only be used for storage and we would convert
to IEEE format for the computation.  Thus our Achilles heal of being
bandwidth limited becomes a great advantage: it provides us a
computational budget to deal with potentially expensive unpacking and
packing at the beginning and end of any computation, respectively.

This paper is broken up as follows: in \S\ref{sec:ieee} we describe
the IEEE numerical formats and briefly introduce the pre-existing QUDA half-precision format that
was a precursor to this work.  In \(\S\ref{sec:solvers}\) we describe the stabilized mixed-precision CG and multi-shift CG solvers that QUDA utilizes, that have not been described in the literature prior. 
 Our alternative numerical formats are described in \S\ref{sec:bitpack}, and we describe the approach to implementing these formats in \(\S\ref{sec:implementation}\), as well as testing the efficacy of our approach.  In \(\S\ref{sec:solver_results}\) we present initial solver results using these numerical formats. Finally in \S\ref{sec:conclusion} we summarize our findings, and we propose
future research directions and applications of our technique.  In this work we present using HISQ staggered fermions, however, there is nothing tied to the discretization employed, and our methods are equally applicable to Wilson fermions, etc.

\vspace{-2mm}
\section{IEEE and QUDA numerical formats}
\label{sec:ieee}
\vspace{-2mm}

The IEEE floating point formats~\cite{8766229} are partitioned into a sign bit, mantissa, and exponent, with the mantissa bits defining the precision of the representation, and the exponent bits setting the possible range.  In Table~\ref{tab:ieee} we summarize the IEEE primary floating point formats.

\begin{table}[htb]
    \centering
    \begin{tabular}{|c|c|c|c|c|c|} \hline
          & exponent & mantissa & epsilon & smallest & largest \\
          & width    & width    &         & number   & number \\ \hline
     half  & 5              & 10 & \(2^{-11}\sim 4.88\times 10^{-4}\) & \(\sim 6.10\times 10^{-5}\) & 65504 \\
     single  & 8              & 23 & \(2^{-24}\sim 5.96\times 10^{-8}\) & \(\sim 1.18\times 10^{-38}\) & \(\sim 3.40\times 10^{38}\)\\
     double  & 11  & 52 & \(2^{-53} \sim 1.11\times 10^{-16}\) & \(\sim 2.23\times 10^{-308}\) & \(\sim 1.80\times 10^{308}\)\\
     quad & 15  & 112 & \(2^{-113} \sim 9.63\times 10^{-35}\) & \(\sim 3.36\times 10^{-4932}\) & \(\sim 1.19\times 10^{4932}\)\\ \hline
    \end{tabular}
    \caption{IEEE 754 floating point formats, together some key properties}
    \label{tab:ieee}
\end{table}
\vspace{-2mm}

A few comments are in order regarding Table~\ref{tab:ieee}.  The quantity epsilon is defined as being the smallest number than can be added to unity, with the result being distinct from unity.  Since floating point errors are relative, this can be thought of as the relative error of a given floating point number.  On the largest and smallest numbers, we note that quantities evaluated on the lattice will typically by \(\mathcal{O}(1)\), with the smallest values typically being \( 1 / \kappa\), where \(\kappa\) here is the condition number of the fermion matrix we are solving for (and not the Wilson mass parameter).  Thus we can observe that it is unlikely we need the huge range afforded by double precision, even if we may need the precision.

Similarly, we may also observe that both the precision and range of IEEE half precision (FP16) are unlikely to be suitable for LQCD given that it would not be possible to represent the dynamic range of a typical linear system to any desirable precision.  It is this that motivates the "half precision" format that is used by QUDA.  This is essentially a fixed-point representation, where we utilize the entire 16-bit effective word size for precision, and do not "waste" any bits for exponent bits.  This is a natural fit for the gauge fields, since these are elements of the special unitary group, and so all values are guaranteed to be in the range \([-1, 1]\).\footnote{The interested reader may wonder what is done in the case of smeared gauge fields, where the matrix elements are no longer necessary bounded as such.  Here we simply compute the global max field element, and use this to set the scale.}  For the fermion fields, since these do not have an {\it a priori} bound, we compute the site-local maximum absolute value and use this to set the scale.  Once being loaded in registers, the fields are converted into FP32 for computation.  Hence the format gives 16 bits of local precision, for an epsilon of \(2^{-15} \sim 3\times 10^{-5}\) while at the same time giving a global dynamic range similar to IEEE single precision.\footnote{One may wonder why we only have precision of \(2^{-15}\), as opposed to \(2^{-16}\): this is because, unlike IEEE floating point formats, we are using a signed integer twos-complement representation, with no separate sign bit.}  Since LQCD computations consists of applying (special) unitary rotations at the site level, a local fixed-point format is in some sense optimal.  In Listing \ref{list:stag_half} we illustrate the structure of the half-precision format as deployed for a staggered fermion in QUDA.


\begin{lstlisting}[language=C++, caption=QUDA half-precision format for a staggered fermion, label={list:stag_half}
]
struct staggered_fermion_16 {
  int16_t v[6];
  float max;
};
\end{lstlisting}

To avoid any confusion, from now on whenever we refer to half precision, we are referring to the QUDA half precision format and not IEEE half precision.  Conversely, single and double precision do refer to their usual IEEE formats.


\vspace{-2mm}
\section{Mixed-Precision Solvers}
\label{sec:solvers}
\vspace{-2mm}

In prior work~\cite{Clark:2009wm}, we reported on an efficient mixed-precision BiCGStab method, where {\it reliable updates}~\cite{reliable} were utilized to ensure that the iterative residual computed in the working, or {\it sloppy}, precision (typically half or single) was periodically replaced with an explicitly computed residual in the target precision (typically double).\footnote{How often the true residual should be computed is a free parameter.  For this work here we use \(\delta = 0.1\), e.g., each time the iterated residual drops by an order of magnitude we recompute the true residual.  We are presently experimenting with a more dynamic choice based on explicit error estimation which may prove optimal\,\cite{replacement}.}  While the reliable-update mixed-precision method could also be applied naively to CG with some limited success, this method was found to break down as the fermion mass was decreased, e.g., condition number increases.  Moreover, mixed-precision methods have additional challenges when one is concerned with multi-shift solvers.  Since the solutions QUDA employs to both of these problems have not previously been reported in the literature, we do so here.  

\vspace{-1mm}
\subsection{Mixed-Precision Conjugate Gradient}
\vspace{-1mm}

An analysis of the breakdown of mixed-precision CG at small quark mass found the issues were centred around a few key defects, which we now describe, together with their respective solution.
\begin{enumerate}
    \item Accumulation of the partial solution vector in the sloppy precision leads to unnecessary truncation of the solution vector.  This can easily be remedied by always storing the solution vector in high precision.  For example, when using a double-half CG solver, while the iterated residual and gradient vectors are kept in half precision, the solution vector is stored in double precision. We can mitigate the additional memory traffic overhead from storing the solution vector in high precision by buffering the gradient vectors, and only accumulating onto the solution vector periodically.
    \item While frequent correction of the residual vector with a reliable update cures the residual drift, it does not prevent drift of the gradient vector.  This means that the super-linear acceleration factor of CG versus Steepest Descent is lost, and the iteration count increases.  This observation regarding the gradient vector drifting was noted in \cite{4020914} with a cure being given to re-project the gradient vector whenever the residual is corrected.
    \item The computation of \(\beta\), the coefficient used to ensure the gradient vectors are A-orthogonal, breaks down when the Krylov space is constructed in low precision.  This has a similar effect as 2.) above, in that the super-linear convergence of CG is lost.  We stabilize the computation of \(\beta\) by adopting the Polak–Ribière formula from non-linear CG~\cite{polak},
\begin{equation*}
    \beta_{k+1} = \frac{r^\dagger_{k+1} (r_{k+1} - r_k)}{r_k^\dagger r_k}.
\end{equation*}
    Note that in infinite precision, the additional term on the numerator disappears since the residual vectors will be orthogonal, however, in finite precision this additional term enhances the local orthonormality.
\end{enumerate}

\begin{figure}[htb]
\includegraphics[width=70mm]{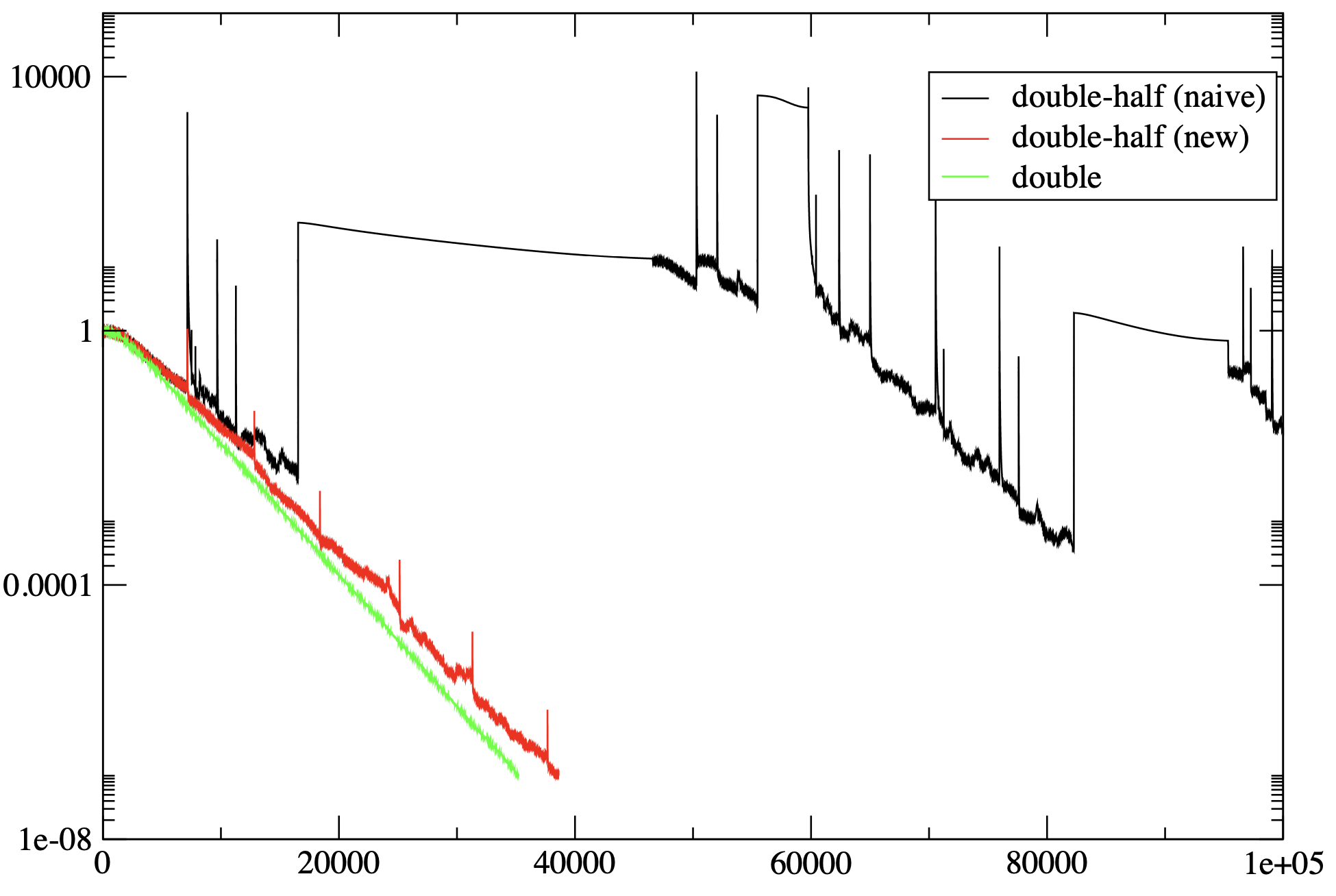}
\caption{Convergence history of CG solver (\(V = 16^4\), \(\beta = 5.6\), \(m = 0.001\)).}
\label{fig:cg}
\end{figure}

In Figure \ref{fig:cg} we plot the residual history of the QUDA CG solver when solving the staggered operator at a relatively light fermion mass.  Shown on the plot are the residual histories for a pure double-precision solver, a naive double-half solver utilizing only reliable updates, and a double-half solver utilizing the above three refinements.  Dramatic improvement is evident, demonstrating that these additions are critical to having a solver that converges at all.  The periodic jumps in the residual are indicative of a reliable update taking place where the residual is recomputed in double precision: that these jumps are mild for the stabilized algorithm indicates that the residual drift is relatively well bounded.

\vspace{-1mm}
\subsection{Mixed-Precision Multi-shift Conjugate Gradient}
\label{sec:multi-shift}
\vspace{-1mm}

Multi-shift solvers rely on the shifted residuals being co-linear to the unshifted base residual throughout the convergence~\cite{https://doi.org/10.48550/arxiv.hep-lat/9612014}, and this proves troublesome when used in combination with mixed precision.  While reliable updates prevent drift of the unshifted residual, they do not prevent drift of the shifted residuals.  Thus over the course of the convergence, the shifted residuals will cease to be co-linear with the unshifted residual, and the shifted systems convergence stalls, even while the unshifted system retains its convergence.

We are chiefly concerned with multi-shift CG, owing to its importance as the primary solver when using Rational Hybrid Monte Carlo (RHMC)~\cite{Clark_2007}.  Prior to this work, the optimal approach employed in QUDA has been to run the multi-shift solver in double-single precision, employing reliable updates on the unshifted system.  As the iterated shifted residuals converge, or reach the limit of the sloppy precision,  we cease updating them and the solver is terminated once all the {\it shifted} systems are removed.  At this point we then employ CG refinement, typically using double-half precision, solving the unshifted system first, then refinement on the shifts.  For the unshifted, we preserve the gradient vector from the initial multi-shift solver, to retain the Krylov space generated during the initial solve, and use a clean restart for the shifted systems.  Compared to a pure double-precision multi-shift solver, we typically achieve around a 1.5-2\(\times\) time-to-solution improvement, with the speedup being tempered by the additional iterations required due to the post-solve shifted system refinement.

\vspace{-1mm}
\subsection{The need for more precision}
\vspace{-1mm}

Even with the techniques and strategies described above there are limits to how far one can push limited precision that cannot be so easily circumvented.
\begin{itemize}
\item When the inverse of the condition number is smaller than the epsilon of the precision employed, any linear solver will struggle since it will be unable to simultaneously represent all eigenvalue scales to even a single significant digit.
\item For solvers that require explicit orthogonalization, e.g., Arnoldi-based methods, the size of Krylov space that can be constructed will be limited by the precision of the representation.
\item As noted above, lack of precision in a mixed-precision multi-shift solve results in a loss of co-linearity.  Improving the sloppy precision could better retain co-linearity in multi-shift solvers, reducing post-solve refinement required, in turn maximizing the benefit of using mixed-precision solvers.
\end{itemize}

Thus we have motivation to consider how to improve the underlying precision of the representation, or to {\it maximize the precision at constant bit width}.

\section{Bit Packing for Lattice QCD}
\label{sec:bitpack}
\vspace{-2mm}

The idea of bit packing to achieve a reduced footprint is not a new one.  A recent work of note in the realm of HPC is ZFP~\cite{6876024} which is a lossy compression library aimed at increasing the precision per bit achievable by using a block storage format and utilizing the spatial coherence that scientific data sets typically exhibit to achieve compression.

In the present case, let us reconsider the staggered fermion half precision format (Listing \ref{list:stag_half}).  The primary value of the FP32 maximum component is the 8-bit exponent, as the 23 bits of mantissa and 1 bit sign add no value relative to the other 16-bit components. A better approach is thus only storing the exponent of the maximum value, and repurpose the remaining 24 bits to improve the precision for all components.  We sketch out such a storage format in Listing \ref{list:stag_20}, where we note we are abusing the C++ type naming convention here to illustrate our point (there are no such types as \texttt{int20\_t} or \texttt{int30\_t}).  With this alternative format we have improved the precision of the fermion components from 16 bits to 20 bits without any increase in memory footprint, which should result in an epsilon of around \(2^{-19} \sim 2\times 10^{-6}\).

\begin{lstlisting}[language=C++, caption=20-bit precision format for a staggered fermion, label={list:stag_20}
]
struct staggered_fermion_20 {
  int20_t v[6];
  uint8_t exponent;
};
\end{lstlisting}

Let us now consider the conventional single-precision format, which for a staggered fermion would consist of six FP32 numbers, or 192-bits.  Here we again can use a shared exponent for all components, and repurpose the saved bits for improving the effective precision.  Listing \ref{list:stag_30} is a realization of this, where we have improved the effective precision to 30 bits (from 24 bits), which should result in an epsilon of around \(2^{-29} \sim 2\times 10^{-9}\).

\begin{lstlisting}[language=C++, caption=30-bit precision format for a staggered fermion, label={list:stag_30}
]
struct staggered_fermion_30 {
  int30_t v[6];
  uint8_t exponent;
};
\end{lstlisting}

Before we discuss implementation of these ideas and results, we note that the above two example formats are just two of the many possible shared-exponent formats that can be constructed: different solutions can be tailored as needs change.  Regarding the gauge field: owing to its lack of global dynamic range we need no such special consideration.  To accompany the 20-bit fermion we simply utilize the 16-bit raw integer format used for half precision and for the 30-bit format we utilize a 32-bit fixed point format.  

Finally we note that a computation involving fermions and gauge fields will require three precisions to be fully specified, and that any one of these can be the precision limiter in a given computation:
\begin{itemize}
    \item The storage format of the gauge field, e.g., 16-bit, 32-bit, FP32
    \item The storage format of the fermion field, e.g., 20-bit, 30-bit, FP32
    \item The actual precision that the computation is done in, e.g., FP32, FP64
\end{itemize}
In the discussion below \(\S\ref{sec:implementation}\), we use the triplet shorthand of gauge precision / fermion precision / computation precision, e.g., 16-bit / 16-bit / FP32 is the conventional QUDA half precision format, and 16-bit / 20-bit / FP32 would be the new format using 20-bit fermions.

\section{Implementation and Testing}
\label{sec:implementation}
\vspace{-2mm}

\subsection{Implementation}
\vspace{-1mm}
While the ideas sketched out in the prior section are easy enough to describe, the fact that there are no native 20-bit and 30-bit integer types in C++ may suggest that implementing such a format in an application framework such as QUDA would be onerous, involving a lot of manual bit masking and shifting.  Fortunately, we can utilize bit fields~\cite{bitfield} to easily address arbitrary subsets of 32-bit words, effectively allowing for arbitrary precision types.  The notable exception here is in the case where we are required to partition a given \(n\)-bit number between multiple (32-bit) words.  For the interested reader, in Appendix \ref{sec:appendix} we have included an example code listing which performs the packing and unpacking of a staggered 20-bit fermion vector.  

By virtue of the fact that QUDA uses opaque data {\it accessors}, providing an abstraction layer between data order and the computation that uses this data, none of these details leak into the high-level computation code: deploying these bit-packed data types is merely a change in template instantiation.

\vspace{-1mm}
\subsection{Precision}
\vspace{-1mm}

\begin{figure}[htb]
\begin{floatrow}
\ffigbox{%
\includegraphics[width=70mm]{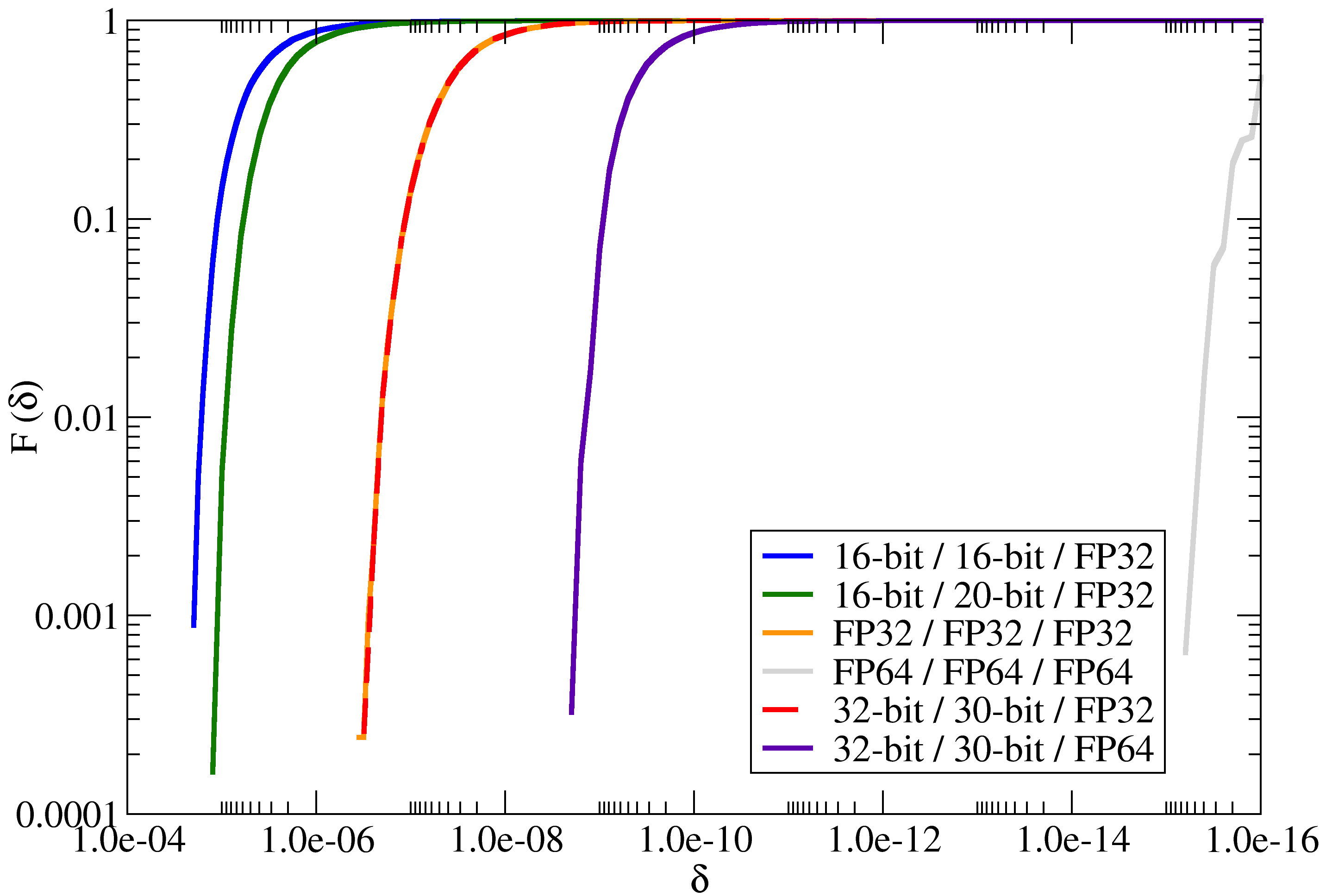}
}{%
    \caption{CDF of the absolute deviation for the application of the HISQ operator to a random source vector (\(V = 36^3\times 72\), \(\beta = 6.3\), \(m = 0.001\))).}
\label{fig:cdf}
}
\ffigbox{%
\includegraphics[width=70mm]{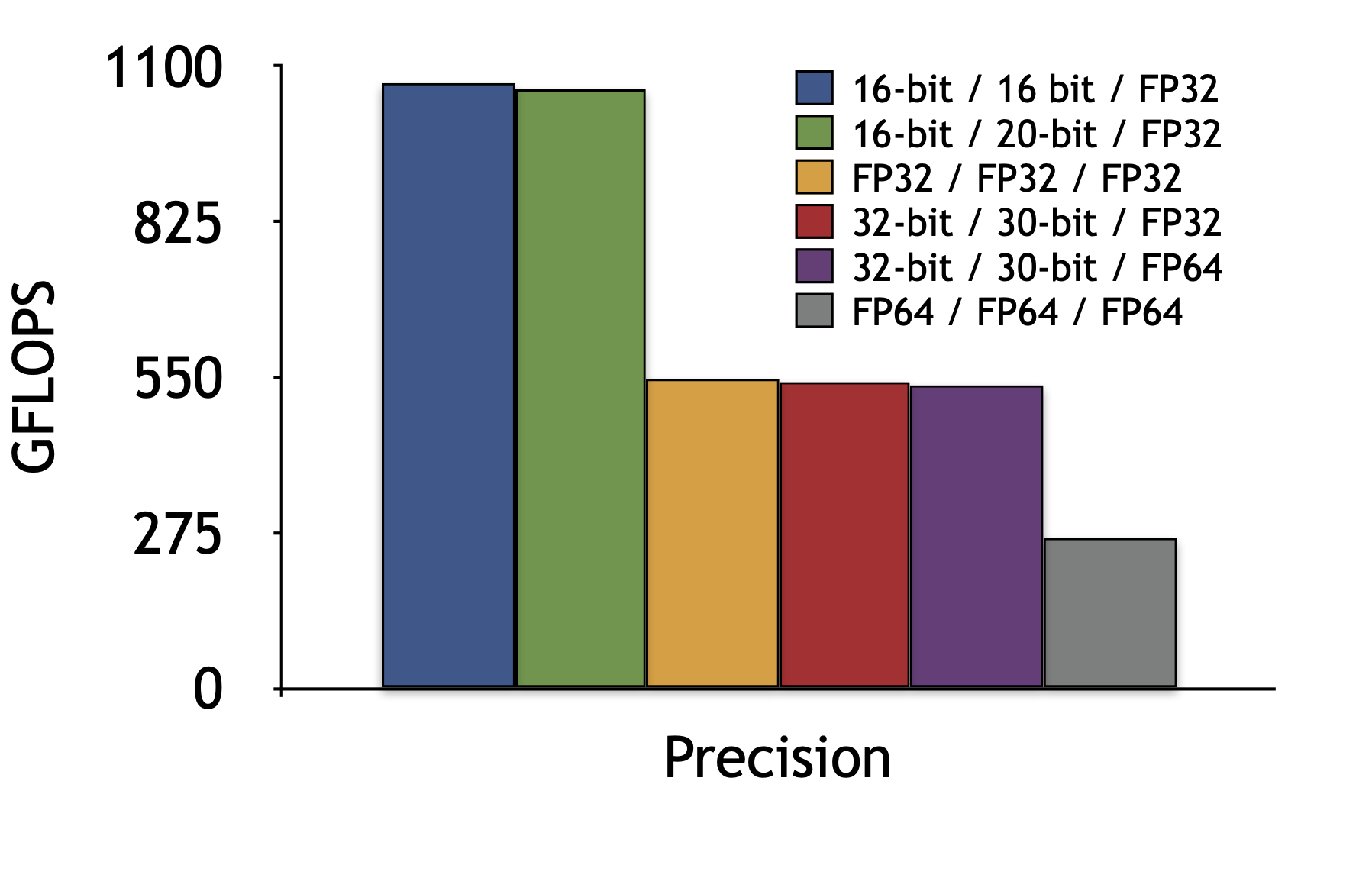}
}{%
    \caption{Performance in GFLOPS for the application of the HISQ operator on a Quadro GV100 (\(V = 32^4\)).}
\label{fig:gflops}
}
\end{floatrow}
\end{figure}

In Figure \ref{fig:cdf} we plot the cumulative distribution function of the absolute deviation of the HISQ Dslash, running on an NVIDIA GPU, using as reference a distinct FP64 implementation running on a host CPU.  For the 20-bit format we see only a modest improvement versus half precision: this is not surprising since in this case we are still using a 16-bit gauge field.  For 30-bit, we see that we must perform the computation using FP64, else we see effectively unchanged precision versus a pure FP32 implementation.  Notably, we see that the 30-bit format, together with FP64 computation achieves around two orders of magnitude additional precision versus pure FP32, with errors \(\sim 10^{-9}\) as we would expect from our epsilon expectations for the 30-bit format.

\vspace{-1mm}
\subsection{Performance}
\vspace{-1mm}

Clearly there is a non-trivial overhead from using these bit-packed data types compared with using the usual IEEE floating point formats.  Thus, even if these formats provide significantly more precision, it may not prove to be a net gain.  In Figure \ref{fig:gflops} we plot the performance of the QUDA HISQ dslash on an NVIDIA GPU.  We see the desired behaviour, that the overhead of using these bit-packed formats is negligible, and performance is essentially identical to the equivalent legacy formats.  It is noteworthy that with the 30-bit format, we can afford the additional packing / unpacking, as well performing all computation using FP64, and yet still have the same performance as pure FP32.  Of course, this statement will be an architecture-dependent one: different architectures can have differing ratios of floating point to integer throughput, and the significant increase in integer processing requirements for unpack and packing these formats could prove significant on other architectures.

\section{Solver Results}
\label{sec:solver_results}
\vspace{-2mm}

In this section we give some examples that illustrate the improved convergence possible from using these bit-packed formats.  For brevity we refer to the 16-bit / 20-bit / FP32 format simply as "20-bit" and similarly the 32-bit / 30-bit / FP64 format as "30-bit".  The results here are illustrative, not definitive, and for this initial technology demonstration we use only a single gauge configuration taken from the NERSC medium benchmark set~\cite{nersc}. 

\vspace{-1mm}
\subsection{BiCGStab($l$)}
\vspace{-1mm}

\begin{figure}[htb]
\begin{floatrow}
\ffigbox{%
\includegraphics[width=70mm]{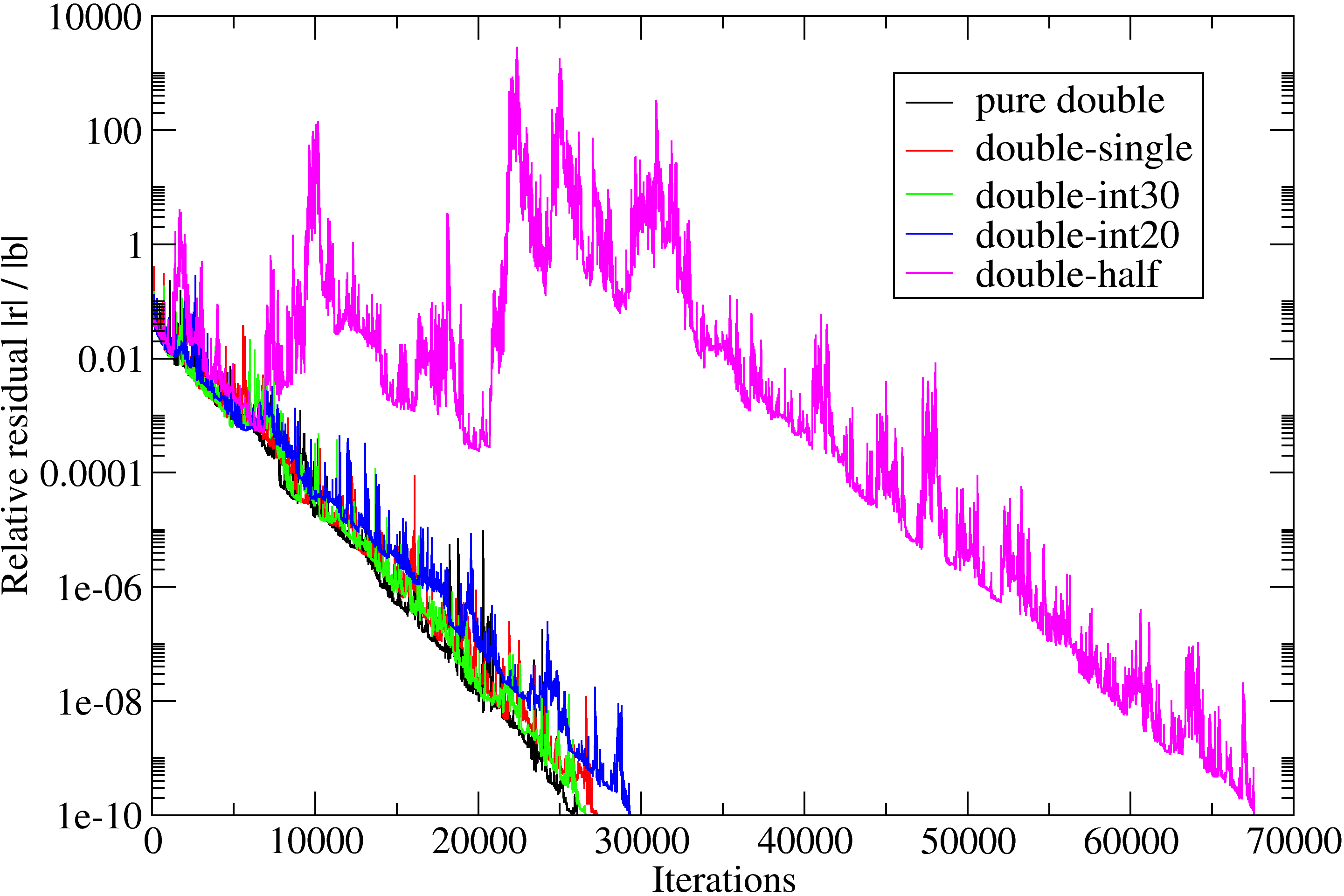}
}{%
    \caption{Convergence history of BiCGStab(4) solver for the HISQ operator (\(V = 36^3\times 72\), \(\beta = 6.3\), \(m = 0.001\))).}
\label{fig:bicg-results}
}
\capbtabbox{%
  \begin{tabular}{|l|c|c|} \hline
                & iter & Time (s) \\ \hline
  pure double   & 26064 & 307 \\
  double-single & 27308 & 159 \\
  double-int30  & 26580 & 150 \\
  double-int20  & 29336 & 106 \\
  double-half   & 67552 & 247 \\ \hline
  \end{tabular}
}{%
  \caption{Iterations and time to solution for the results shown in Figure \ref{fig:bicg-results}.}%
\label{tab:bicg-results}
}
\end{floatrow}
\end{figure}

The BiCGStab(\(l\)) solver is an enhancement over the BiCGStab solver in that it extends the single iteration MR minimization to \(l\) iterations of GCR minimization~\cite{sleijpen1993bicgstab}.  Thus each \(l\) steps we are required to orthogonalize an \(l\)-length vector set.  In principle a greater value of \(l\) will lead to smoother convergence, however, lack of precision in the orthogonalization can result in solver breakdown.  In Figure \ref{fig:bicg-results} we plot the convergence history from using BiCGStab(4) for the HISQ staggered operator using pure double, double-single, double-int30, double-int20, and double-half mixed-precision solvers.  The corresponding time to solution and iteration counts are shown in Table \ref{tab:bicg-results}.  We see improved convergence for double-int30 versus double-single resulting in a reduced time to solution.  The stand out improvement comes from double-int20 versus double-half, where we see the improved precision from the vectors (fermions) alone dramatically stabilizes the solver convergence leading to the optimal approach.  We attribute this improvement to the increased stability of orthogonalization in the solver.

\vspace{-1mm}
\subsection{Multi-shift CG}
\vspace{-1mm}

For the baseline  we deploy the mixed-precision strategy as described in \(\S\)\ref{sec:multi-shift}, and compare this to an improved variant where we run the initial multi-shift solver in double-int30 (as opposed to double-single) and the refinement CG solver now uses double-int20 (opposed to double-half).  In Table \ref{tab:multi-shift} we compare the iteration counts for an example multi-shift CG solve.\footnote{This is taken from a light quark solve in RHMC, each shift has a unique target residual here where the tolerance is set to optimize for the resulting force contribution\,\cite{https://doi.org/10.48550/arxiv.hep-lat/0510004}.}  For both approaches we see it is the intermediate shifts that require the greatest refinement iterations: this is because they are far enough away from the unshifted system that their residuals have drifted sufficiently, while at the same time not being heavy enough to have fast convergence.  For the unshifted system we see virtually identical convergence for new versus old.  However, when we look at the shifted systems we see in all cases more than an order of magnitude reduction in the residual.  As a result we see a dramatic reduction in the additional refinement iterations required, going from hundreds of iterations required per shift to a few dozen per shift.  The overall effect is a reduction in total number of operator applications by around 13\%.

\begin{table}
  \begin{tabular}{|l|c|c||c|c||c|} \hline
   & \multicolumn{2}{c||}{double-single-half} & \multicolumn{2}{c||}{double-int30-int20} & target residual \\ \hline
  shift & iter & \(r_\sigma\) & iter & \(r_\sigma\) & \\ \hline
  (multi-shift)& 8728 & - & 8726 & - & 1.00e-05\\\hline
  0     & 2323 & 8.91e-05 & 2320 & 8.92e-05 & 1.00e-05\\\hline
  1     &   40 & 4.82e-03 & 10 & 2.62e-04 & 1.75e-05 \\
  2     &  113 & 1.82e-03 & 12 & 9.93e-05 & 3.86e-05 \\
  3     &  432 & 6.14e-04 & 15 & 3.33e-05 & 8.85e-05 \\
  4     &  411 & 1.95e-04 & 18 & 1.06e-05 & 2.05e-08\\
  5     &  323 & 6.08e-05 & 22 & 3.29e-06 & 4.81e-09 \\
  6     &  232 & 1.87e-05 & 44 & 1.01e-06 & 1.15e-09 \\
  7     &  158 & 5.64e-06 & 60 & 3.01e-07 & 2.50e-10 \\
  8     &  114 & 2.03e-06 & 56 & 1.04e-07 & 4.64e-11 \\
  9     &   81 & 9.10e-07 & 45 & 4.24e-08 & 1.30e-11 \\
  10    &   64 & 5.96e-07 & 37 & 2.41e-08 & 6.85e-12 \\ \hline
  total & 13019&          & 11365 & &\\ \hline
  \end{tabular}
  \caption{Multi-shift and subsequent refinement CG solver iteration counts: shift 0 represents a continuation of the multi-shift solve but on the unshifted system only.  For each shift we show the true residual after the initial multi-shift solve \(r_\sigma\) and the target residual (\(V = 36^3\times 72\), \(\beta = 6.3\), \(m = 0.001\)).}%
\label{tab:multi-shift}
\end{table}

\section{Conclusion}
\label{sec:conclusion}
\vspace{-2mm}

In this work we have introduced LQCD-specific storage formats that achieve more precision at fixed bit width compared to the conventional IEEE floating point formats.  Given the bandwidth-bound nature of LQCD computations, this method can be an important optimization approach.  The effect on mixed-precision solvers was illustrated, where we observe we can reduce the total number of iterations at fixed bit width without any significant raw performance impact.

This work is just an initial investigation and there a number of possible applications for these techniques.  Beyond solver stabilization, one application could be offline storage using more efficient bit-packed formats for storing gauge fields or propagators.  One interesting application may be for extending beyond double precision: present-day LQCD computations are starting to bump up against the precision limits of IEEE double precision, with some gauge generation implementations now selectively utilizing 128-bit precision floating point precision~\cite{openqcd}, whether it be true quad or double-double.   As the need for extended precision increases, an alternative would be to use a Tailor-made precision format with just the {\it right} amount of precision for the application at hand. 

\clearpage

\begin{appendices}

\section{Example: 20-bit staggered fermion vector}
\label{sec:appendix}

We include the C++ code below as an example of how to implement the packing and unpacking for a 20-bit staggered fermion.  One implementation detail we call out: while we have signed numbers in general, bit shifting should be done with unsigned integers to prevent undefined behaviour.  

\begin{lstlisting}[language=C++, caption=20-bit format for a staggered fermion, label={list:packer_stag_20}
]
  template <unsigned int B, typename T> T signextend(const T x) {
    struct {T x:B;} s;
    s.x = x;
    return s.x;
  }
  
  union float_structure {
    float f;
    struct float32 {
      unsigned int mantissa : 23;
      unsigned int exponent : 8;
      unsigned int sign     : 1;
    } s;
  };
  
  template <> struct alignas(16) spinor_packed<20> {
    static constexpr unsigned int bitwidth = 20;
    static constexpr float scale = 524287; // 2^19 - 1

    // the struct consists of 4x 32 bit words
    unsigned int a_re : 20;
    unsigned int a_im_hi : 12;

    unsigned int a_im_lo : 8;
    unsigned int b_re : 20;
    unsigned int b_im_hi : 4;

    unsigned int b_im_lo : 16;
    unsigned int c_re_hi : 16;

    unsigned int c_re_lo : 4;
    unsigned int c_im : 20;
    unsigned int exponent : 8;

    // Pack a 3 component complex vector into this
    template <typename spinor> void pack(const spinor &in) {
      // find the max
      float max = {fabsf(in[0].real()), fabsf(in[0].imag())};
      for (int i = 1; i < 3; i++) {
        max = fmaxf(max, fabsf(in[i].real()));
        max = fmaxf(max, fabsf(in[i].imag()));
      }

      // compute rounded up exponent for rescaling
      float_structure fs;
      fs.f = max[0] / scale;
      fs.s.exponent++;
      fs.s.mantissa = 0;
      exponent = fs.s.exponent;

      // negate the exponent to avoid the division below
      fs.s.exponent = -(fs.s.exponent - 127) + 127;

      // rescale and convert to integer
      int vs[6];
      for (int i = 0; i < 3; i++) {
        vs[2 * i + 0] = lrint(in[i].real() * fs.f);
        vs[2 * i + 1] = lrint(in[i].imag() * fs.f);
      }

      unsigned int vu[6];
      for (int i = 0; i < 6; i++) memcpy(vu + i, vs + i, sizeof(int));

      // split into required bitfields
      a_re = vu[0];
      a_im_hi = vu[1] >> 8;

      a_im_lo = vu[1] & 255;
      b_re = vu[2];
      b_im_hi = vu[3] >> 16;

      b_im_lo = vu[3] & 65535;
      c_re_hi = vu[4] >> 4;

      c_re_lo = vu[4] & 15;
      c_im = vu[5];
    }

    // Unpack into a 3-component complex vector
    template <typename spinor> void unpack(spinor &v) {
      // reconstruct 20-bit numbers
      unsigned int vu[6];
      vu[0] = a_re;
      vu[1] = (a_im_hi << 8) + a_im_lo;
      vu[2] = b_re;
      vu[3] = (b_im_hi << 16) + b_im_lo;
      vu[4] = (c_re_hi << 4) + c_re_lo;
      vu[5] = c_im;

      // convert to signed
      int vs[6];
      for (int i = 0; i < 6; i++) memcpy(vs + i, vu + i, sizeof(int));

      // signed extend to 32 bits and rescale
      float_structure fs;
      fs.f = 0;
      fs.s.exponent = exponent;

      using real = decltype(v[0].real());
      for (int i = 0; i < 3; i++) {
        v[i].real(static_cast<real>(signextend<bitwidth>(vs[2 * i + 0])) * fs.f);
        v[i].imag(static_cast<real>(signextend<bitwidth>(vs[2 * i + 1])) * fs.f);
      }
    }
  };
\end{lstlisting}

\end{appendices}

\bibliographystyle{JHEP}
\bibliography{lattice_2022}

\end{document}